\documentclass[aps,preprintnumbers,amsmath,amssymb,superscriptaddress,floatfix,nofootinbib,a4paper]{jpconf}

\usepackage{lipsum}
\usepackage{graphicx}
\usepackage{epsfig}
\usepackage{epstopdf}
\usepackage{hyperref}
\usepackage{amsmath}
\usepackage{amsfonts}
\usepackage{amssymb}

\usepackage{subfig}
\usepackage{slashed}
\usepackage{soul}
\usepackage[normalem]{ulem}
\usepackage{float}
\usepackage{subfig}
\usepackage{epsfig}
\usepackage{pstricks}
\usepackage{color}
\usepackage{mathrsfs}

\def\ve{\varepsilon}
\def\ver{\varepsilon_R}

\def\beqa{\begin{eqnarray*}}
\def\eeqa{\end{eqnarray*}}

\def\hlf{\frac{1}{2}}

\def\cblack{\color{black}}

\def\sc{\scriptsize}

\def\vsp{\noalign{\vskip 0.3cm}}

\def\be{\begin{equation}}
\def\ee{\end{equation}}
\def\bea{\begin{eqnarray}}
\def\eea{\end{eqnarray}}
\def\beas{\begin{eqnarray*}}
\def\eeas{\end{eqnarray*}}
\def\bfr{\begin{flushright}}
\def\efr{\end{flushright}}
\def\bfl{\begin{flushleft}}
\def\efl{\end{flushleft}}
\def\begenu{\begin{enumerate}}
\def\endenu{\end{enumerate}}
\def\bit{\begin{itemize}}
\def\eit{\end{itemize}}
\def\bdes{\begin{description}}
\def\edes{\end{description}}
\def\bqu{\begin{quote}}
\def\equ{\end{quote}}
\def\hlf{{1\over 2}}

\def\delW{\delta Z_W^{1/2}}
\def\delA{\delta Z_{AA}^{1/2}}
\def\delZA{\delta Z_{ZA}^{1/2}}
\def\delAZ{\delta Z_{AZ}^{1/2}}

\def\rar{\rightarrow}

\def\xnla{\tilde{\alpha}}
\def\xnlb{\tilde{\beta}}
\def\xnlk{\tilde{\kappa}}
\def\xnld{\tilde{\delta}}
\def\xnle{\tilde{\varepsilon}}

\def\ve{\varepsilon}

\def\nn{\nonumber}

\begin{document}
\title{Automatic calculation of two-loop ELWK corrections to the muon  ($g$-2)${}^*$}

\author{Tadashi Ishikawa$^1$, Nobuya Nakazawa$^2$ and  Yoshiaki Yasui$^3$}

\address{$^1$ High Energy Accelerator Organization(KEK), 1-1 OHO Tsukuba Ibaraki 305-0801,Japan}
\address{$^2$ Department of Physics, Kogakuin University, Shinjuku,Tokyo 163-8677,Japan}
\address{$^3$ Department of Management, Tokyo Management College,Ichikawa, Chiba 272-0001, Japan}
\ead{nobuya@suchix.kek.jp}

\begin{abstract}
 Two-loop electroweak corrections to the muon anomalous magnetic moment
are automatically calculated by using GRACE-FORM system, 
as a trial to extend our system for two-loop calculation.
We adopt the non-linear gauge (NLG) to check the reliability of our calculation.
In total 1780 two-loop diagrams consisting of 14 different topological types and 70
one-loop diagrams composed of  counter terms are calculated. We check UV- and 
IR-divergences cancellation and 
the independence of  the results from NLG parameters.
As for the numerical calculation, 
we adopt trapezoidal rule with Double Exponential method (DE). 
Linear extrapolation method (LE) is 
introduced to regularize UV- and IR- divergences and to get finite values.
\end{abstract}

\section{Introduction}

In order to get a sign of beyond the standard model physics from high precision experimental data, 
 we need  higher order radiative corrections within Standard Model~(SM). For this purpose our group
 has been developing the  automatic calculation system GRACE \cite{GRACE} since the late 1980's.
 The measurement of the muon anomalous magnetic moment ($g$-2) is 
 the one of the most precise experiments to 
 check the SM. QED correction was calculated by T.Kinoshita et al. \cite{Kinoshita} up to tenth-order.
 The two-loop electroweak (ELWK) correction to ($g$-2) was calculated approximately by  
 Kukhto et al. \cite{Kukhto} in 1992.
 Surprisingly,  the two-loop correction is almost 20\% of the one-loop correction. 
 We started to calculate the full two-loop corrections in 1995 and presented
 our formalism at Pisa conference \cite{Pisa}.  We also showed that  
 the two-loop QED value \cite{Karplus} was correctly reproduced within our general formalism. 
 However, the number of diagrams is huge (1780+70) and the numerical integration
 requires the big CPU-power to achieve required accuracy, 
 we must wait until various environments are improved. 
 
 During these days, the several groups did the calculations using leading log($M^2$)
 approximations. ($M$=heavy particle mass) \cite{Marciano} \cite{Grib} 
 and the approximate value of the two-loop ELWK correction is widely accepted \cite{Marquard}.
  In 2001, BNL-Experiment 821 \cite{BNL1} announced that the precise experimental 
 value deviates from that of SM around $(2.2 \sim 2.7)~\sigma$.
It brought  much interest in the theoretical value. \\
\hrulefill
\hrule width 16cm
\bfl
${}^*$\footnotesize{To appear in the proceedings of the 4th Computational Particle Physics Workshop, October 2016, Hayama, Japan.}
\efl
\newpage
The various  theoretical predictions and experimental value are summarized in the next subsection.
 The main theoretical concern is shifted to the hadronic contributions.
 \subsection{present status of the theoretical predictions of ($g$-2) and future experiments}
 We summarize the present theoretical predictions of muon ($g$-2) in Table \ref{thvalue}.

\begin{center}
\lineup
\begin{table}[h]
\caption{\label{thvalue}Theoretical predictions of muon ($g$-2)~~\cite{PDG} } 
\begin{tabular}{@{}l*{15}{r}{l}{c}}
\br  
Type of correction &Numerical Value (unit $10^{-11}$) &Error& Reference \\
\mr
QED up to tenth-order &116584718.95 & (0.08) & \cite{Kinoshita} \\
Leading Order Hadronic Vac.Pol. & 6923~~~~& (42)(3)&\cite{Davier}   \\
NLO Vac.Pol.+Hadronic LBL &  7\0\0&(26) & \cite{Prades}  \\
ELWK (one-loop)& 195.82& (0.02) & \cite{Weak1} \\
ELWK (two-loop)&  - 41.2\0& (1.0) & \cite{Kukhto}\cite{Marciano}\cite{Grib}\cite{Weak2} \\
\hline
Theory total & 116591803~~~~& (1)(42)(26) & \cite{PDG}    \\
\hline
Experimental Value & 116592091~~~~&(54)(33)& \cite{BNL1}\cite{CODATA}\\
\br
\end{tabular}
\end{table}
\end{center}
The discrepancy between the experimental value and 
the theoretical value is still large. As new experiments are 
scheduled at FNAL-E989 \cite{FNAL} and J-PARC-E034
\cite{Jparc}, we can expect to have new data within a few years.

\subsection{Purpose of  our calculation}
Although the two-loop ELWK correction is almost established
as  summarized  in the Table \ref{thvalue}, 
we try to get the value without any approximation to confirm 
the validity of the earlier studies. 
It is also an important trial to extend GRACE-system 
from one-loop to  two-loop calculation.  It may become a
milestone to construct the framework of  
Perturbative Numerical Quantum Field Theory (PNQFT).
The concepts of PNQFT  are  summarized as follows.
(a) It is essential to assume  amplitudes as meromorphic functions
of space time dimension $n$, for extracting both UV- and IR-divergences.
 By adopting dimensional regularization,  gauge invariant  
 renormalized value is obtained directly.
(b) The source for numerical integration is  automatically generated 
by a symbolic manipulation system.
(c) It is crucial to reduce human intervention to avoid careless mistakes,
so that numerical method is fully exploited.   
The following calculation shows the outline of this direction.

In section 2 and 3, we briefly explain the flow and foundation of our 
calculation.
In section 4, we touch on our method of numerical calculation.
We emphasize that the Linear Extrapolation (LE) method \cite{LEM} is simple and efficient method
to regularize UV- and IR-divergences and also to get finite values.
In numerical calculation, the validity of the results 
must be guaranteed by comparing several independent methods.
We explain shortly our consistency conditions
 to ensure the results.  
We also show some examples of 
calculations and how the consistency conditions are satisfied.
In section 5, we give a part of our results on ($g$-2) for restricted types 
of diagrams. 
In the last section, we summarize the present status and
give some comments to make extensive progress.
\section{Outline of our frame work}
Our calculation is formulated under the following conditions.
\begenu
\item
Adopt non-linear gauge (NLG) formulation with 't Hooft-Feynman propagator.
\item
Dimensional regularization for both Ultra Violet (UV)- and Infrared (IR) -divergences.
\item
On mass shell renormalization scheme is adopted.
\item
Linear Extrapolation method (LE) is fully used for regularization and 
getting finite values.\\
~~
\endenu
Next, we briefly explain the flow of our calculation.
\begenu
\item
GRACE system generates all the diagrams we need in SM.
There are 1780 two-loop diagrams and 70 one-loop diagrams
composed of  one-loop order counter term (CT).
\item
These 1780 diagrams are classified into 14 types of topology.
Types of  the topology are displayed 
in Fig.1. 
Among these types, some of them give the same contribution
because of  symmetry. (an example: 5-a vs. 5-b )
 The diagrams including CT
are essentially two types, namely, vertex and self-energy types.
\endenu
\makeatletter
\newcommand{\figcaption}[1]{\def\@captype{figure}\caption{#1}}
\newcommand{\tblcaption}[1]{\def\@captype{table}\caption{#1}}
 \begin{center}
\includegraphics[scale=0.30]{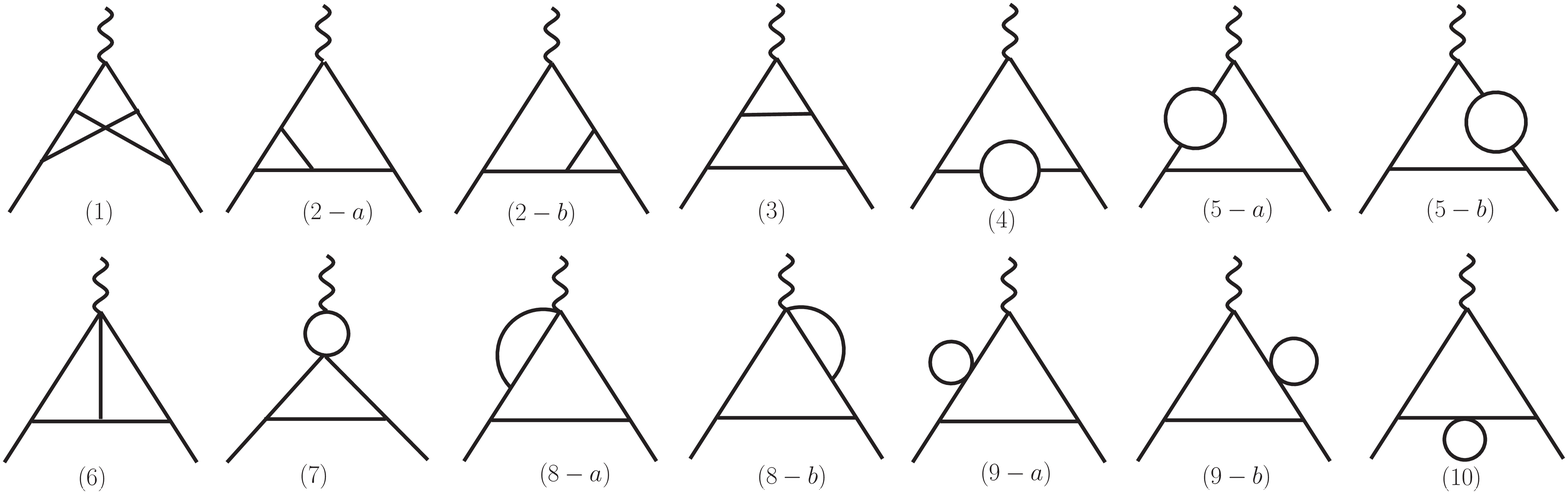}

\figcaption{Types of  topology}
 \label{topol}
 \end{center}
\begenu
\setcounter{enumi}{2}
\item
In order to distinguish the UV-part from
 finite part, we prepare a few files 
determined by the type of topology, in advance.
They represent the following quantities. \cite{Civita}
\bit
\item
Internal loop momentum flow$(\ell_s,\eta(s))$($s=1,2$)
\item
External momentum flow $(q_j)$   
($j=$ internal line number)
\item
Kirchhoff's law of momentum conservation at each vertex
\eit
\item
Using these tools,  both UV-divergent and finite contributions
to ($g$-2) are calculated in turn, according to the formulas given in the 
next section~\cite{Civita}.
We make use of a symbolic manipulation system FORM \cite{FORM}
exhaustively.
\endenu
\section{Logic of the calculation}
We explain the case where there are six-propagators.
Starting formula is the two-loop muon vertex,
\bea
\Gamma_\mu 
= \int
 \frac{d^n \ell_1}{i(2\pi)^n}
 \frac{d^n \ell_2}{i(2\pi)^n}
 \frac{F_\mu(D)}{\prod_j (p_j^2-m_j^2)}
 =\Gamma(6)\int
 \prod dz_j \delta(1-\sum_j z_j)
 \int
 \frac{d^n \ell_1}{i(2\pi)^n}
 \frac{d^n \ell_2}{i(2\pi)^n}
 \frac{F_\mu
 (D)}{\sum_j z_j(p_j^2-m_j^2)}
\eea
 where $\displaystyle{p_j=\sum_{s=1}^2\eta(s)\ell_s +q_j}$, is the  momentum on the internal line $(j)$.
The function  $\eta(s)(=\pm 1 ,0) $  defines loop momentum 
on the internal line $(j)$. 
The $z_j$'s are the Feynman parameters to combine six propagators.
The numerator function $F_\mu(D)$  is explained later. 

Next  we diagonalize the denominator function with respect to loop 
momenta $\ell_1$ and  $\ell_2$ and perform 
integration. The result is,
\bea
\Gamma_\mu 
=\frac{1}{(4\pi)^n}\int
 \prod dz_j \delta(1-\sum_j z_j)
\frac{\Gamma(6-n)}{({\rm detU})^{n/2}}
F_\mu(D)\frac{1}{(V-i\varepsilon)^{6-n}},~~
U_{s,t}=\sum_{j=1}^6z_j\eta_s(j)\eta_t(j)  \label{IR0}.
\eea
Where ${U}$ is well known 2$\times$2 matrix, composed of 
Feynman parameters $(z_j)$.  
$V(z_j,m_j, q_j)$ is the denominator function. 
The arguments are easily 
extended to the case with five-propagators (diagrams with four-point coupling).  

In order to generate the numerator function we use  the following 
differential integral operator $D_\mu$.
\bea
\frac{p_j^\mu}{(p_j^2 - m_j^2)}
=D_j^\mu \frac{1}{(p_j^2 - m_j^2)},
~~~~{\rm where }~~~
D_j^\mu \equiv  \frac{1}{2}\int\nolimits_{m_j^2}^{\infty}
dm_j^2 \frac{\partial}{\partial q_{j\mu}}
\eea
The operator $D_j^\mu$ generates momentum $p_j^\mu$ on the internal line $(j)$.
By operating $D_j^\mu$ to the denominator function $V$, 
we get the following expression.
\bea
D_j^\mu\frac{1}{V^m} =\frac{Q'_{j\mu}}{V^m},
~~~Q'_j = q_j -\frac{1}{\rm detU} \sum_{i=1} z_iB_{ij}q_j,
~~~B_{ij}=\sum_{s,t}\eta_s(i)\eta_t(j)U^{-1}_{st}{\rm detU} = B_{ji}
\eea
Similarly we can generate the higher order term.
\bea
D_i^\mu D_j^\nu \frac{1}{V^m}
=\frac{Q'_{i\mu}Q'_{j\nu}}{V^m}
+\left(- \frac{1}{2\rm detU}\right)\frac{g_{\mu\nu}}{(m-1)}
       \frac{B_{ij}}{V^{m-1}} \label{2ndD}  \label{TKformula}
\eea
We can also derive higher order terms such as 
$D_i^\mu D_j^\nu D_k^\lambda(1/V^m)$  and so on, however, 
Eq.(\ref{2ndD}) is sufficient in our case. 
We can verify the equivalence of the above method and
the well known method of shifting momentum in the numerators.
The correspondence between two methods are symbolized as follows.
\bea 
\ell^0~\rar~ \frac{\{1,Q'_{i\mu},Q'_{i,\mu}Q'_{j,\nu},\cdots\}}{V^m},~~~~\ell_i\ell_j~~\rar~~ 
\left( -\frac{1}{2{\rm detU}}\right)\frac{g^{\mu\nu}}{(m-1)}\frac{B_{ij}}{V^{m-1}}
\eea

Next step is to extract the ($g$-2) factor by using projection operator,
from the photon muon vertex $\Gamma_\mu$. 
The quantity ($g$-2) is given as follows. ( $m_0$=muon mass )
\bea
(g{\rm -2}) &=& \lim_{q^2\rightarrow 0} \frac{{m_0}}{p^4q^2}
 {\rm Tr}\left(\Gamma_\mu {\rm Proj }(\mu) \right)  \nn \\
 \vsp
{\rm Proj}(\mu) &=& \frac{1}{4}(\rlap{/}p-\hlf\rlap{/}q+m_0)
    \{{\rm m_0}\gamma_\mu(p.p)-({m_0}^2+\frac{q.q}{2})p_\mu \}
    (\rlap{/}p+\hlf\rlap{/}q + {\rm m_0}),~~~
\eea
where we set momentum of incoming $\mu^-$, outgoing $\mu^+$  and incoming photon, as 
$(p-q/2),(p+q/2)$ and $q$, respectively.  

Next step is the regularization of UV-divergence.  By adopting $n$-dimensional
regularization method,
any integrand $F$ of Feynman parameter integration is regarded as a function of  $\ve=4-n/2$. 
Starting from the function $F(\ve)$, we adopt two methods for regularization.
The first method is well known one and  extract $1/\ve$ singularity from $F(\ve)$ when
one of Feynman parameters  approaches  0,
($x\rightarrow 0$). Another one is called LE-method  explained below \cite{LEM}. We compare two methods
and conclude LE-method is simpler and more efficient than the first one if the accuracy of  numerical integration 
is increased considerably.

First we explain the first method.  
In order to extract UV-divergence,  we use  the following procedure.
First we transform the Feynman parameters $(z_1,z_2,\cdots,z_6)$ into the appropriate [0,1] variables
$(x,y,u,v,w)$ depending on the topology. The key point is to factorize the function detU=$x\times u(x,\cdots)$,
where $u(0,\cdots)\ne 0$ . Then it is shown that the UV-part in dimensional regularization
comes from the following formula. We show the example of vertex type renormalization.
We set $n=4-2\ve,~~\ve>0$ for UV-regularization.
\bea
I=\int_0^1dx x^{\ve-1}F(x,\ve)=\frac{1}{\ve}F(0,0)+\frac{\partial F(0,0)}{\partial \ve}
+ \int_0^1\frac{F(x,0)-F(0,0)}{x} dx  \label{vertex-type}
\eea
In the case where there is self-energy type diagram, we need the formula 
$ \int_0^1 dx x^{\ve-2}F(x,\ve)$ in addition to Eq.(\ref{vertex-type}). So it becomes more complex. 

The second method named LE  is applicable for both UV- and IR-regularization and 
obtaining finite values \cite{LEM}.
Space time dimension $n$ is set to be $(4-2\ve)$ for UV-case or $(4+2\ver)$
for IR-case. In both cases, ~$\ve>0,  \ver >0$. 

 After  dimensional regularization method was introduced \cite{n-dim}, the analyticity with respect to $\ve$ was discussed extensively. 
 It is shown that the Feynman amplitude is a meromorphic function of $\ve$ \cite{n-dim} \cite{Nakanishi}. 
This is a key point to utilize the LE-method to the Feynman amplitude. 
 Numerically, we calculate $G(i) =\int F(z_j,\ve)\prod dz_j$  when $\ve=\ve(i)$.
$(i=1,2,\cdots M)$. We set  $\ve(i)=1/\alpha^{i+14}$ by taking relevant value $\alpha$.
According to the analyticity, we can expand $G$ and truncate  it at  O($\ve^{M-2}$).
\bea
G(i) = c_{-1}\frac{1}{\ve(i)}+c_0 + c_1\ve(i)+\cdots +c_{M-2}\ve(i)^{M-2}=\sum_{j=-1}^{M-2}c_j\{\ve(i)\}^j
~~~~(i=1,2,\cdots M)
\eea
From this formula, we get $\{c_{j}\}$ by multiplying  the inverse of  $M\times M$ matrix $A$, 
whose element is  $A(i,j)=\{\ve(i)\}^j,~(i=1,\cdots M,~j=-1,0,\cdots M-2)$, to  M-component vector $G(i)$. 
Practically we set $M\sim15$ and $ \alpha \sim 1.15$. 

As for counter terms, GRACE has a library of renormalization constants at one-loop level. We make use of
this library for 70-diagrams composed of  counter terms. 
We use  FORM \cite{FORM} to generate  Fortran source, according to the above formulation.
\section{Numerical Calculation}

The final step to get the value ($g$-2) is the numerical integration over Feynman parameters.
We employed trapezoidal rule with Double Exponential (DE) transformation method  \cite{HTakahashi}.
It is very powerful if the integrand has singular behavior at the edge of  the integration domain.
Speed of convergence is accelerated by the DE transformation,
\bea
I=\int_{0}^{1}dx f(x)~\rar~ x = \phi(t)=\hlf
\left\{1+\tanh\left(\frac{\pi}{2}\sinh(t)\right) \right\}
\eea
The maximum dimension of multiple integration is five. We apply DE-method to
any integration parameter involved. 
As we need the accuracy  of 7$\sim$10   digits to see  the cancellation of  UV,  the
adaptive Monte Carlo method is not suitable for two-loop calculation.

In order to ensure the validity of our results, we impose several conditions given below.
\begenu
\item
Well known QED two-loop value is reproduced.
\item
UV-divergence is cancelled.
\item
IR-divergence is cancelled.
\item
The result is independent of Non-linear gauge parameters.
\item
In some cases (examples: topology 4,5-a,5-b,7,9-1,9b and 10) ,
we perform loop-integrations $(\ell_1,\ell_2)$ in turn,
 (we call it successive method) and obtain the same results.
\endenu
In all these cases, if we have plural methods to evaluate, we compare the numerical values 
to ascertain the validity. If all these five conditions are cleared, we can insist on the 
validity of our calculation.

We demonstrate how the conditions are cleared by showing the examples.
\bit
\item
\underline{\bf Reproduction of QED} two-loop value. (Unit =$ (\alpha/\pi)^2 )$ \\
Analytic expression     -0.328478966  ~~~~~: our value    -0.328478911 
\eit
\bit
\item
\underline{\bf UV-cancellation} in Feynman gauge (FG) among the diagrams. \\
Examples of a group of diagrams are shown in Fig.2.
\eit
In this case, total 13-diagrams and 1-counter term
make a group to cancel UV-divergence. As you can read from the table,
the cancellation is marvelous, up to almost 10 digits.

 \begin{center}
 \includegraphics[scale=0.35]{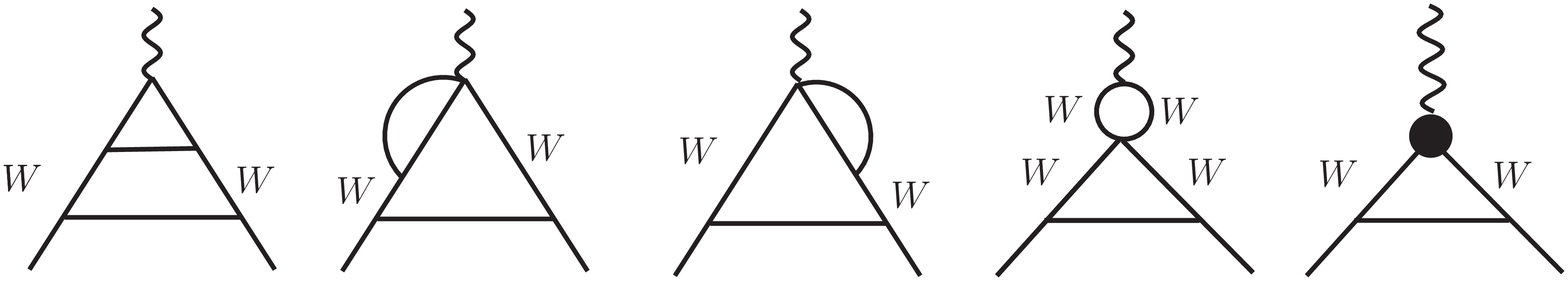}
 \label{UVcancel}
 \figcaption{Sample of diagrams with $W$}
 \end{center}

\makeatletter
 \begin{center}
\tblcaption{Sample of UV-cancellation in FG}
    \label{left1}
\begin{tabular}{ccr}
\br
Topology & Diagram No.&   Value ~~$( \rm{unit~ }10^{-11})$\\
\mr
 3&149 & 1.679393199868\\
3&153 &  0.490572808433 \\
3&157 &   0.055640154059\\
3&161 & 0.055640154059\\
3&171 & -0.021530682050\\
3&172 & -0.006289394980\\
3&173 & -0.021530682050\\
3&174 & -0.006289394980\\
8-a&1517 & -0.166093750630\\
8-a&1518 & -0.048518165797\\
8-b&1578 & -0.166093750630\\
8-b&1579 & -0.048518165797\\
7& 1671 & -2.575344486333\\
CT-vtx &1781 & 0.778962156811\\
\br
 & Sum & -0.000000000017\\
\end{tabular}
    \end{center}
\bit
\item
\underline{\bf IR-cancellation} is also checked by using LE-method.
Among two-loop diagrams,  the 8 diagrams in Fig.3 have IR-divergence.
The diagrams with CT also have IR-divergence through $\delW$ (19-diagrams) and 
$\delta Z_\mu^{1/2}$ (28-diagrams).
 In LE-method, the IR-divergence is proportional to 
$C_{IR}^{(2)}= (-1/\ver -2\gamma+2\ln(4\pi))$,  $\ver=(n/2-2)$. 
It is easily shown that the IR-divergence coming  from $\delW$ cancels 
among the 19-CT-diagrams. As for the diagrams with $\delta Z_\mu^{1/2}$,
IR-divergence is cancelled by the corresponding two-loop diagrams. 
 We show  coefficients  of $C_{IR}^{(2)} $ in Table \ref{IRcan}.
 The correspondence between small photon mass ($\lambda$) method and LE-method for IR-regularization
is checked in the case of QED ladder diagram.
Analytic value of a coefficient of  $\ln(\lambda^2/m_\mu^2)$
in unit of $(\alpha/\pi)^2$ is $(1/4)$  \cite{Karplus}. It is  0.249999998 by our calculation 
using small photon mass. 
In LE-method, the coefficient of $C_{IR}^{(2)}$ becomes  0.249999999.  We understand 
the correspondence between  $\ln(\lambda^2/m_\mu^2)$ and  $C_{IR}^{(2)}$  is established.
\begin{center}
  \includegraphics[scale=0.25]{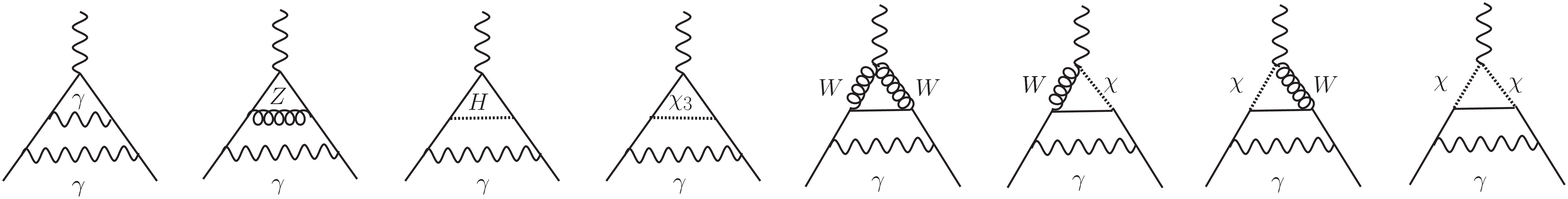}
 \label{IRcancelfig}
 \figcaption{IR cancellation among diagrams with $W$ and $\chi^{\pm}$ bosons}
 \end{center}
\begin{center}
\tblcaption{Cancellation of IR-divergence~~(unit $10^{-11}$)}
\label{IRcan}
\begingroup
\renewcommand{\arraystretch}{1.2}
\scalebox{0.9}{
\begin{tabular}[htbp]{ccccc}
\br
Diagram Type & $\delta Z_\mu^{1/2}$ Vertex CT &  $\delta Z_\mu^{1/2}$ Self CT & two-loop diagram &sum  \\
 \mr
 Neutral Type  &    &   &  diagram  &  \\
    $\gamma$  &  $-134887.2756 \times 3$ & $134887.2756 \times 2$  &  (a) $+134887.2755$ & $-1.6 \times 10^{-4}$ \\
    $ Z-boson$   & $0.2152351626 \times 3$ & $-0.2152351626 \times 2 $ &(b) -0.2152351628 & $-1.7 \times 10^{-10}$  \\
    $ Higgs, \chi_3$ & $1.66293755 \times 10^{-6} \times 3$ & $-1.66293755 \times 10^{-6} \times 2$ & 
$-1.66293757 \times 10^{-6}$  & $-2.5 \times 10^{-14}$ \\
\mr
 Charged type &      &    &   diagram(c)$\sim$(h) & \\
   $ W^{\pm} ,\chi^{\pm}$ & -0.4312728113299& $---$ &0.4312728113330    & $3.1\times 10^{-12}$\\
\br
\end{tabular}
}
\endgroup
\end{center}

As we  show in Table \ref{IRcan},  no IR-divergence remains in the final expression.
\item
\underline{\bf Non-linear Gauge (NLG)} parameter independence is also checked. Non linear gauge was introduced
to reduce the number of diagrams, particularly containing boson-boson couplings \cite{Boudjema}.
Here, we adopt NLG to check the validity of our calculation.
The gauge fixing Lagrangian is constructed as,
\bea
\mathscr{L}_{GF}=-\frac{1}{\xi_W}F^+F^- -\frac{1}{2\xi_Z}(F^Z)^2
-\frac{1}{\xi}(F^A)^2
\eea
where
\bea
F^{\pm}&=&\left( \partial^\mu\mp ie \xnla A^\mu \mp i\frac{e c_W}{s_W}\xnlb Z^\mu \right)W_\mu^\pm
+\xi_W\left ( M_W\chi^\pm+\frac{e}{2s_W}\xnld H\chi^\pm 
\pm i\frac{e}{2s_W}\xnlk \chi_3\chi^\pm     \right) \nn \\
F^Z&=&\partial^\mu Z_\mu + \xi_z\left( M_Z \chi_3 + \frac{e}{2s_Wc_W}\xnle H\chi_3\right) \nn \\
F^A&=&\partial^\mu A_\mu
\eea
Here, $\xnla,\xnlb,\xnld,\xnle$ and $\xnlk$ are non-linear gauge parameters specific to 
this formalism. The parameters $s_W$ and $c_W$ are the $sine$ and $cosine$  of Weinberg angle $\theta_W$. 
In our calculation we set $\xi=\xi_W=\xi_Z=1$ to make the gauge boson propagators 
simple.
As an example of  cancellation of NLG parameters, we show the UV-divergent part proportional
to $\xnla^2$. The 53 diagrams make a group and we show sum of a coefficient of 
$C_{UV}^{(2)}=(1/\ve-2\gamma+2\ln(4\pi))$ for the group of the same topological type in the 
Table \ref{UV-linear}.
 By summing up these contributions, 
12 digits cancellation is confirmed even in 
4 to 5 dimensional integration over Feynman parameters. \\
 \begin{center}
\tblcaption{UV-part~~($\xnla^2$-term)}
    \label{UV-linear}
\begin{tabular}{ccr}
\br
 Type  &  Number of & Coeff. of $C^{(2)}_{UV}$\\
    & diagrams &   in unit $(\alpha/\pi)^2$        \\
\mr
2a &   4 & 5.994906222E-08 \\
2b & 4  & 5.994906222E-08 \\
3 & 7   & -9.325381915E-06\\
5a &13&1.398809747E-07 \\
5b & 13 &1.398809747E-07 \\
8a &  2 & 4.422894730E-06  \\
8b & 2  & 4.422894730E-06 \\
9a & 4 & 3.996606459E-08 \\
9b &  4  &3.996606459E-08 \\
\mr
sum(*)  & 53 & -6.048741686E-18 \\
\br
\end{tabular}
\end{center}

{\footnotesize (*) Each contribution  is calculated in quadruple precision 
method and has more effective digit than shown in the table.} 
\item
\underline{\bf Successive method} is also applied to two-loop diagrams with 
self energy type two-point function. An example is diagram with 
$(\gamma-\gamma)$ or $(\gamma-Z)$ vacuum polarization type diagram.
We decompose the renormalization constants $\delA,\delZA,\delAZ$ etc.
 into components according to the particles
involved in the loop. By adding the corresponding counter term
 to one loop unrenormalized two-point function, we 
integrate over the first loop momentum $\ell_1$. 
The result is  finite renormalized two-point function  $\Pi_R(\ell_2)$. 
The function $\Pi_R$ contains the second loop momentum $\ell_2$,
however, the divergent part of $\ell_2$ integration does not 
contribute to ($g$-2). We use this alternative method to 
reconfirm the results obtained  by the methods given in section 3.
\eit
\section{Finite contribution from restricted type of diagrams}
The publicly accepted value of ELWK two-loop correction
is classified  according to  types of diagrams.  For example,  fermion loop 
contribution with or without Higgs boson and  purely bosonic contribution 
are discussed separately. 
As an example, we summarize the naive fermion loop contribution
without any approximation. The diagrams we consider are shown in 
Fig.4.
\begin{center}
 \includegraphics[scale=0.30]{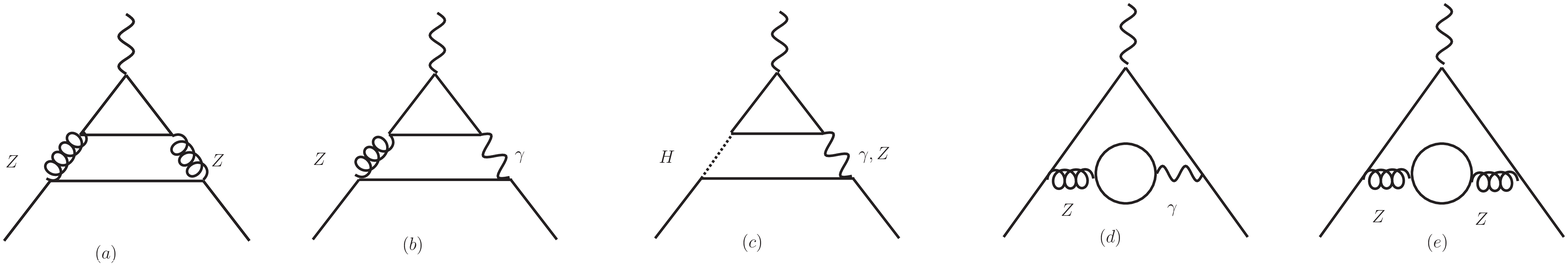}
 \label{Fermion}
 \figcaption{Fermion loop diagrams}
 \end{center}
 
As is shown in Fig.4, we consider diagrams with both fermion loop and neutral boson.

The contribution from fermion triangle diagrams Fig.(4-a,b) comes from VVA-anomaly. 
If  lepton and  quark mass  within the same generation are degenerate, VVA-anomaly cancellation
works so that the contribution becomes zero. So fermion mass difference is crucial to get the 
value. The $\ln(M_z^2)$-dependence cancels within one generation, because of no-anomaly condition.
 Czarnecki et al.\cite{Marciano} give the approximate formula for diagram 
Fig.(4-b). Taking the same parameters (see below), we compare their values and our exact results
in Table \ref{CMours}.
\begin{table}[htb]
\caption{\label{CMours} Comparison of Fig.(4-b)~~~~~~~~~~~~(unit $10^{-11}$)}
\begin{center}
\begin{tabular}{lcc}
\br
lepton and quark &  Ref.\cite{Marciano} approximate value& Our exact value\\
\mr
electron, up,down    &-3.9984&-4.0272\\
muon, charm,strange  &-4.6524&-4.6426\\
$\tau$, top, bottom  &-8.1854&-8.1940\\
\mr
Total               & -16.8362 ~~& -16.8638~~\\
\br
\end{tabular}
\end{center}
\end{table}

Although each fermion contribution is quite different, 
the coincidence becomes pretty good by taking the sum in each generation.
The difference comes from the common constant terms 
which cancels by summing up within each generation, 
due to the anomaly free condition.
We also show the preliminary value of other diagrams Fig.(4-a,c,d,e) 
in Table \ref{Floop}.
 \begin{center}
\tblcaption{Finite value Fig(4-a,c,d,e)}
    \label{Floop}
\begin{tabular}{ccr}
\br
 Type  &   Value ~~$\{10^{-11}\}$\\
\mr
4-a &-0.0494 \\
4-c &  -1.8718\\
4-d & -0.1803 \\
4-e &  \ 1.5842 \\
\br
\end{tabular}
\end{center}
We use the following fermion and boson masses in the calculation in unit GeV.
$m_{\mu}= 105.658389\times 10^{-3},     
  m_{e} = 0.51099906\times 10^{-3}, m_{\tau}= 1.7771,
  m_u = 0.3,m_c = 1.5,
  m_t = 173.21,m_d = 0.3,
  m_s = 0.5,m_b=4.5, M_W = 80.22,
  M_Z= 91.187, M_H=125.09 $

\section{Discussion and Comment}
We develop the system to calculate the full ELWK two-loop corrections to ($g$-2). 
Unfortunately it is not 
quite completely automatic system yet.  We can produce Fortran source automatically.
 The work we need beforehand is only to prepare several files
 which only depend on the type of the topology of diagrams
 as we explained in section 2.
 
If we adopt the well known first method to extract UV-divergent part, it is crucial to introduce 
the most suitable transformations from $(z_j)$ to [0,1] integration variables $(x,y,u,v,w)$.
In the case of Linear Expansion method (LE), however, 
the choice of  integration variables is not sensitive to get the reliable results.

Further comments are given on the regularization method. Firstly, we introduce $n$-dimensional regularization
method and take out UV-divergence $(1/\ve)$, when one of the integration variables is approaching
0 ($x \rightarrow 0$). In order to perform this procedure, we need rather complex operations 
including differentiation of  the amplitudes, etc.
As a result, it  makes  Fortran source lengthy so that  CPU-time increases extensively.
The second prescription to regularize both UV- and IR-divergences is LE-method,
 briefly explained in section 3.
This method decreases the number of operation drastically. 
It is sufficient to define the quantity as function of $\ve(=2-n/2)$ .
We only need to treat Dirac matrices and various vectors appeared in the 
numerator, in $n$-dimension. 
This is easily done by using symbolic manipulation system such as FORM.
The operation is simple and we can make use of the resultant short sources 
 for both UV-($\ve>0$) and IR-($\ver=-\ve > 0$) regularization and 
 also to get finite results.
 By comparing the results of 
various methods mentioned in the previous sections, we conclude that LE-method is 
the most simple and reliable method, at this moment.
In order to get  reliable physical value by this method, 
accurate numerical integration over Feynman parameters is 
inevitable.  The DE-method introduced in section 4 is the suitable candidate.

By using these technical approaches mentioned above, we clear almost all the constraints given in section 4.
Namely, (i) reproduction of QED values, (ii)(iii) UV-and IR-divergences are cancelled, 
 (iv) the result is independent of NLG-parameters.
 We show some samples how they are cleared. We are now recalculating and
  rechecking the constraints under the best computer circumstances. 
  The final value will be given soon. 
 We expect ongoing work provides the fruitful foundation to formulate PNQFT
  (Perturbative Numerical Quantum Field Theory) mentioned in section 1.2.

\ack
We would like to thank Prof.T.Kaneko for his important contribution to construct the 
framework of  calculation at the early stage of this work. We also wish to 
thank Prof.K.Kato and Prof.F.Yuasa for discussion. Last but not least, we 
express our deep appreciation to late Prof.Y.Shimizu for his continual encouragement.
This research is partially supported by Grant-in-Aid for 
Scientific Research (15H03668,15H03602) of JSPS.


\end{document}